\documentstyle[prl,preprint,aps]{revtex}

\def\AmS{{\protect\the\textfont2
        A\kern-.1667em\lower.5ex\hbox{M}\kern-.125emS}}

\makeatletter
\tighten
\begin{document}
\draft
\title{A Landau theory for the metal-insulator transition}
\author{T.R.Kirkpatrick}
\address{Institute for Physical Science and Technology,\\
University of Maryland,\\
College Park, MD 20742}
\author{D.Belitz}
\address{Department of Physics and Materials Science Institute,\\
University of Oregon,\\
Eugene, OR 97403}

\date{\today}
\maketitle
\begin{abstract}
The nonlinear $\sigma$-model for disordered interacting electrons is studied
in spatial dimensions $d>4$. The critical behavior at the metal-insulator
transition is determined exactly, and found to be that of a standard
Landau-Ginzburg-Wilson
$\phi^4$-theory with the single-particle density of states as
the order parameter. All static exponents have their mean-field values,
and the dynamical exponent $z=3$. $\partial n/
\partial \mu$ is critical with an exponent of $1/2$, and the electrical
conductivity vanishes with an exponent $s=1$. The transition is qualitatively
different from the one found in the same model in a $2+\epsilon$
expansion.
\end{abstract}

\narrowtext

\par
Despite much progress in our understanding of the metal-insulator transition
(MIT) in recent years\cite{R}, the search for a simple order parameter
description of this phase transition in the spirit of a mean-field or
Landau theory has so far proven futile. Early attempts at a mean-field theory
of the Anderson transition of noninteracting electrons
\cite{HarrisLubensky} failed because
the most obvious simple order parameter (OP),
viz. the single-particle density of states (DOS) at the Fermi
level turned out to be uncritical\cite{WegnerDOS}.
As a result, the Anderson transition can be described only in terms of a
complicated functional OP\cite{FyodorovMirlin}.
At the Anderson-Mott transition of disordered interacting electrons,
on the other hand, the
DOS is generally believed to be critical, but it is not obvious how to
construct an OP description in terms of it. Indeed,
our understanding of this transition is
largely based on a generalized matrix nonlinear $\sigma$
(NL$\sigma$)-model\cite{F}, for which no mean-field type fixed point (FP)
is known, and which has been analyzed in terms
of a small disorder expansion near $d=2$. As a result,
there is no simple description of the MIT analogous to, say, Weiss theory of
ferromagnetism. This is all the more remarkable because of the technical
similarity, first noted by Wegner\cite{F}, between the NL$\sigma$-model
description of the MIT and that of Heisenberg ferromagnets.

\par
In this Letter we show that the NL$\sigma$-model for interacting disordered
electrons\cite{F} possesses a saddle point solution which has all of the
characteristic features of a Landau theory with the DOS as the OP. We
further show that this solution corresponds to a renormalization group (RG)
FP which is stable for $d>4$. This establishes the exact (for
$d>4$) critical behavior at the MIT in this model, which can be summarized
as follows:  Let $t$ be the dimensionless distance from the critical point at
temperature $T=0$, $\Omega$ the energy distance from the Fermi level, and
$Q$ the DOS.  $Q$ vanishes according to
\begin{mathletters}
\label{eqs:1}
\begin{equation}
Q(t,\Omega = 0) \sim t^\beta,\ \  Q(t=0, \Omega) \sim \Omega^{\beta/\nu z}  .
\label{eq:1a}
\end{equation}
For the critical exponents $\beta$, $\nu$, and $z$, and for the exponents
$\gamma$ and $\eta$ characterizing the OP susceptibility, we find
\begin{equation}
\nu = \beta = 1/2,\ \  \gamma = 1,\ \  \eta = 0,\ \  z = 3  .
\label{eq:1b}
\end{equation}
\end{mathletters}
All thermodynamic susceptibilities show the same critical behavior,
\begin{equation}
\chi(t, \Omega =0) \sim t^{1/2} ,\ \  \chi(t =0, \Omega) \sim \Omega^{1/3}   ,
\label{eq:2}
\end{equation}
where $\chi$ can stand for the density susceptibility
$\partial n/\partial \mu$, the specific heat coefficient
$\gamma =\lim_{T\rightarrow 0}C_{V}(t,T)/T$, or
the spin suceptibility $\chi_s$.
As an argument of susceptibilities, $\Omega$
denotes the external frequency, and
$\Omega$ and $T$ can be used interchangeably in a scaling sense.
The electrical conductivity $\sigma$ vanishes according to
\begin{equation}
\sigma(t, \Omega = 0) \sim t,\ \  \sigma(t = 0, \Omega) \sim \Omega^{2/3}  ,
\label{eq:3}
\end{equation}
so the conductivity exponent $s=1$.

\par
In what follows we first derive Eqs.(\ref{eqs:1}) by explicitly
constructing the saddle point solution,
and then using RG techniques to show that it is stable for $d>4$.
We then use scaling arguments to obtain additional information about
susceptibilities, which leads to Eqs.(\ref{eq:2}),(\ref{eq:3}).
We consider the matrix NL$\sigma$-model of
Refs.\ \onlinecite{F,R}, i.e. a Gaussian field
theory for a hermitian matrix field $\tilde{Q}({\bf x})$ with constraints
$(\tilde{Q}({\bf x}))^2 = \openone$,
with $\openone$ the unit matrix, and $tr \tilde{Q}({\bf x})=0$. $\tilde{Q}$
is a classical field comprising two fermionic fields. It carries two
Matsubara frequency indices $n,m$ and two replica
indices $\alpha,\beta$ (quenched disorder
has been incorporated by means of the replica trick).  The matrix elements
$\tilde{Q}^{\alpha\beta}_{nm}(\bf{x})$ are in general spin
quaternions, with the quarternion degrees of freedom describing the
particle-hole and particle-particle channel, respectively.
The action\cite{F,R} can be written
\widetext
\begin{eqnarray}
S[\tilde{Q},\Lambda]=\frac{-1}{2G}\int d{\bf x}\ tr
\biggl[\Bigl(\Lambda({\bf x})[\tilde{Q}^2({\bf x})-
\openone]\Bigr)+tr\Bigl(\partial_{\bf x}\tilde{Q}({\bf x})\Bigr)^2\biggr]+2H
\int
d{\bf x}\ tr\Bigl(\Omega \tilde{Q}({\bf x})\Bigr)
\nonumber\\
-\frac{\pi T}{4} \sum_{u=s,t,c} K_u[\tilde{Q}({\bf x})\circ \tilde{Q}({\bf
x})]_u  .
\label{eq:4}
\end{eqnarray}
\narrowtext
Here $G$ is a measure of the disorder, $\Omega$ is a diagonal matrix whose
elements are the external Matsubara frequencies $\omega_n$, and $H$ is
proportional to the free electron DOS.
$K_{s,t,c}$ are coupling constants describing the electron-electron
interaction in the particle-hole spin-singlet,
particle-hole spin-triplet, and
particle-particle or Cooper channel, respectively.
$K_s<0$ for repulsive interactions.
$[\tilde{Q}\circ\tilde{Q}]_{s,t,c}$ denotes a product in
frequency space which is given explicitly in
Refs.\ \onlinecite{F,R}. Notice that we have enforced the constraint
$\tilde{Q}^{2}=\openone$ by means of an auxiliary matrix field
$\Lambda({\bf x})$. The constraint
$tr\ \tilde{Q}=0$ will be enforced explicitly at every stage of the theory.

\par
The model, Eq.(\ref{eq:4}), is an effective one which was
derived to capture the essence of the low-lying excitations of the system,
i.e. the low-frequency large-wavelength behavior. The soft (i.e., diffusive)
modes are given by correlation functions of the $\tilde{Q}_{nm}$ with
$nm<0$, while the DOS is determined by the average of
$\tilde{Q}^{\alpha\alpha}_{nn}$\cite{F,R}. It is therefore convenient to
separate $\tilde{Q}$ into blocks:
\widetext
\begin{equation}
\tilde{Q}^{\alpha\beta}_{nm}=
\Theta(nm)Q^{\alpha\beta}_{nm}({\bf x})+\Theta(n)\Theta(-
m)q^{\alpha\beta}_{nm}({\bf x})+\Theta(-
n)\Theta(m)(q^+)^{\alpha\beta}_{nm}({\bf x}).
\label{eq:5}
\end{equation}
The conventional treatment of the NL$\sigma$-model\cite{F,R}
proceeds by integrating out $\Lambda ({\bf x})$, using
the constraint $\tilde{Q}^2=\openone$ to eliminate $Q$, and expanding
the action in powers of $q$.
Here we use a different approach inspired by treatments of
the O(N)-NL$\sigma$-model in the large-N limit\cite{ZJ}.

\par
Since the further development will be closely
analogous to the treatment of an
O(N)-Heisenberg model, let us pause to point
out the similarities between these models.
The matrix elements $Q_{nm}$ correspond to the massive $\sigma$-component of
the O(N)-vector field, while the $q_{nm}$ correspond to the massless
$\pi$-fields. The disorder $G$ plays the role of the
temperature in the magnetic model, it is the
control parameter for the phase transition.
$H\Omega$ is in some respects analogous to the
magnetic field conjugate to the order parameter $\sigma$.  The last term in
Eq.(\ref{eq:4}) has no analogy in the Heisenberg model.
We will see that in the mean-field treatment of the matrix model presented
here it plays a rather trivial, although crucial, role.

\par
We proceed by integrating out the massless $q$-field.  Since the action,
Eq.(\ref{eq:4}), is quadratic in $\tilde{Q}$ and hence in $q$,
this can be done exactly.  We then look for
a saddle point of the resulting effective action $S[Q,\Lambda]$.
This task is simplified by restricting ourselves
to saddle point solutions that are spatially constant, diagonal matrices
with diagonal elements $Q_n^{\alpha}$, $\Lambda^{\alpha}_n$.
This {\it ansatz} is motivated by the fact that $<Q_{nm}^{\alpha \beta}>$
has these properties.
For simplicity we restrict ourselves to the particle-hole spin-singlet
channel, i.e. we put $K_t=K_c=0$. We will see later that this restriction
does not influence the critical behavior.
We further use a short-range model interaction, so that $K_s$ is
simply a number. Again it can be shown that a Coulomb interaction leads
to the same critical behavior\cite{ustbp}.
We find for the saddle point solution,
\widetext
\begin{mathletters}
\label{eqs:6}
\begin{equation}
(Q^{\alpha}_n)^2=1+ \frac{G}{4}\sum_{\bf p}\sum^{-{\infty}}_{m=-1}\frac{2\pi
TGK_s}{[p^2+\frac{1}{2}(\Lambda^{\alpha}_n +\Lambda^{\alpha}_m)]^2}
\left[1+\sum^{n-m-1}_{n_{1}=0}\frac{2 \pi TGK_s}{p^2 +
\frac{1}{2}(\Lambda^{\alpha}_{n_{1}} + \Lambda^{\alpha}_{n_{1} - n+m})}
\right]^{-1}
\label{eq:6a}
\end{equation}
\begin{equation}
\Lambda^{\alpha}_n = 2GH \omega_n/Q^{\alpha}_n   .
\label{eq:6b}
\end{equation}
\end{mathletters}
\narrowtext
We discuss several aspects of this result.  Firstly,
$Q^{\alpha}_{n = 0}\equiv Q$, which is the DOS normalized
by the free electron result, decreases
with increasing disorder (remember $K_{s}<0$).
This is the well known 'Coulomb anomaly' in the
DOS\cite{AA}, and it is proportional to $K_s$ as well as $G$.
Technically, this is due to the replica structure of the theory:
All terms on the r.h.s. of Eq.(\ref{eq:6a}) that are
independent of $K_s$ vanish in the replica limit and have not been shown.
Physically, it reflects that fact that the Coulomb anomaly is due to the
{\it interplay} between interactions and disorder.
Secondly, Eqs.(\ref{eqs:6})
constitute an integral equation for $Q^{\alpha}_n$ which
involves an integration
over all frequencies.  As noted above, the NL$\sigma$-model is
designed to describe
only low-frequency behavior,
and can not be trusted at high frequencies.  However,
on physical grounds it is clear that $Q^{\alpha}_n$ tends to a constant at
large frequencies, so for fixed $K_s$ $Q$ will vanish at a
critical value $G_c$ of $G$. We can easily determine the
critical behavior for $\delta G \equiv G-G_c<0$:
We find $Q\sim (-\delta G)^{1/2}$, which is the first relation in
Eq.(\ref{eq:1a}).  Furthermore, the leading low-frequency behavior
for $G=G_c$ involves only integrals up to
the external frequency, which is the
region where the theory is controlled. For $d>4$ we find $Q(\Omega)\sim
\Omega^{1/3}$, which is the second relation in Eq.
(\ref{eq:1a}) with the exponent
values given in Eq.\ (\ref{eq:1b}).

\par
We have shown that our saddle point actually corresponds to a minimum of
the free energy by expanding the action to second order in the
fluctuations $\delta Q, \delta \Lambda$ about the saddle point.  Details of
this
calculation will be reported elsewhere\cite{ustbp}.  The result is a positive
definite
Gaussian matrix, so the saddle point is indeed a minimum.  One can then
integrate
out $\delta \Lambda$ to obtain the order parameter correlation function
$G_{n_1n_2,n_2n_4}({\bf k})$.  At zero frequency, and close to the critical
point
(i.e., for $Q \rightarrow 0$) the result simplifies substantially and we find
\begin{equation}
G_{00,00}({\bf k})=\frac{G/4}{k^2+Q^2/\xi_{0}^2}    ,
\label{eq:7}
\end{equation}
where the bare correlation length
$\xi_{0}$ is given in terms of a complicated integral
which is finite for $d>4$. From Eq.(\ref{eq:7})
we obtain three more critical exponents:  $\nu=\beta=1/2, \eta=0$, and
$\gamma=2\nu=1$, cf. Eq.(\ref{eq:1b}).

\par
Apart from the DOS we are interested in the transport properties. Let us
first consider the charge diffusion coefficient $D_{c}$. As a hydrodynamic
quantity it can be obtained from the NL$\sigma$-model by a direct calculation
of the particle-hole spin-singlet $q$-propagator\cite{F,R}. Inserting the
saddle point solution, Eqs.(\ref{eqs:6}), into Eq.(\ref{eq:4}), one
reads off the $q$-vertex function, and a matrix inversion yields the
corresponding propagator. The latter has the usual diffusion pole structure,
and $D_{c}$ is obtained as the coefficient of the momentum squared. We find
\begin{equation}
D_{c} = \frac{Q}{G(H+K_{s}Q)}\ \ .
\label{eq:Dc}
\end{equation}
We see that at the transition $D_{c}$ vanishes like the OP. The behavior of
the conductivity can be obtained by multiplying $D_{c}$ with
$\partial n/\partial \mu$. The latter is less straightforward to obtain, since
as a thermodynamic quantity it is not simply given by the hydrodynamic
behavior. We will determine its critical behavior from Eqs.(\ref{eqs:10})
below.

\par
We now use RG techniques to show that these results represent the
{\it exact} critical behavior of the model in $d>4$.  The RG will also
provide an easier route to a derivation of
Eqs.(\ref{eq:2}) and (\ref{eq:3}) than a direct calculation would be.
Let us return to the action,
Eq.(\ref{eq:4}).  If we integrate out $q$, we get Eq.({\ref{eq:4}) with
$\tilde{Q}$ replaced by $Q$ plus terms obtained by contracting $q$-fields.
Of the latter, one term is linear in $\Lambda$.
The resulting action can be written in the form
\widetext
\begin{eqnarray}
S[Q, \Lambda] = -c \int d{\bf x}({\bf \partial_{\bf x}}Q({\bf x}))^2
- \int d{\bf x}\  tr(t \Lambda ({\bf x}))
+ 2H \int d{\bf x}\  tr (\Omega Q({\bf x}))\nonumber\\
-\frac{\pi T}{4}
\sum_{u=s,t,c} K_u[Q({\bf x})\cdot Q({\bf x})]_u+u_1 \int d{\bf x}\ tr
(\Lambda({\bf x})Q^2({\bf x}))
\nonumber\\
+ \int d{\bf x} \sum_{I,J} \Lambda_I({\bf
x})(u_2)_{IJ}\Lambda_J({\bf x})+ \rm{(other \ terms)}  .
\label{eq:8}
\end{eqnarray}
\narrowtext
Here $c=u_1=1/2G$ and $t$ is a matrix composed of $-(1/2G)\openone$
and the term of $O(\Lambda)$ coming from the $q$-contractions. $I\equiv(n
m,\alpha\beta,i)$ where $i$ labels the spin-quaternion space, and
$u_2$ is a matrix which is finite for $d>4$.  The explicit expressions for $t$
and
$u_2$ will not be needed for our present purposes,
they will be given elsewhere\cite{ustbp}.
The 'other terms' in Eq.
(\ref{eq:8}) all come from contracting $q$-fields.  They can easily be
constructed diagrammatically, but will turn out to be irrelevant for our
purpose.

\par
We now apply standard power counting to the action,
Eq.(\ref{eq:8}).  Our parameter space is spanned by
$\mu=\{c,t,H,K_{s,t,c},u_1,u_2\}$, and the coupling constants of
the 'other terms.'  In looking for a RG fixed point (FP), we follow
Refs.\ \onlinecite{MaFisher} and \onlinecite{ZJ} in fixing the
exact scale dimensions of our fields $Q$ and $\Lambda$ to be $[Q]=d/2-1$,
$[\Lambda]=d-2$ (we define the scale dimension of a length to be -1).  This
corresponds to fixing the exponents $\eta$ and $\tilde{\eta}$ defined by the
wavenumber dependence of the $Q$- and $\Lambda$-correlation functions to be
$\eta = 0$ and $\tilde{\eta}=d-4$, respectively.  With $[\Omega]=[T]=d$, we
find the scale dimensions of the various coupling in
Eq.(\ref{eq:8}) to be $[c]=0, [t]=2, [H]=1-d/2,
[K_{s,t,c}]=2-d, [u_1]=[u_2]=4-d$.  A power counting analysis of all of the
'other terms' in
Eq.(\ref{eq:8}) is straightforward.  The result is that the coupling
constants of all
of these terms have scale dimensions which are smaller than $4-d$ for $d>4$.
We conclude that the Gaussian FP given by $\mu^{*}=\{c,0,0,0,...\}$
is stable for $d>4$. The only relevant parameter is $t$, and
the correlation length exponent $\nu$ is $\nu = 1/[t]=1/2$.
c is marginal, as
expected, and all other parameters, including H and $K_{s,t,c,}$, are
irrelevant (the
combinations $H\Omega$ and $K_{s,t,c}T$ are, of course, relevant, which
reflects the fact that a finite frequency or wavenumber
drives the system away from the
critical point). This Gaussian FP obviously corresponds to the saddle point
solution
discussed above. It also is, structurally, just the ordinary Gaussian FP of
$\phi^4$-theory:  For scaling purposes, $\Lambda$ is equivalent to $Q^2$.
This is
a consequence of our choice of the scale dimensions of Q and $\Lambda$, and
becomes explicit if one neglects the irrelevant 'other terms' in
Eq.(\ref{eq:8}) and integrates
out $\Lambda$.  The stability of the FP proves that the critical behavior
obtained from the saddle point solution is exact in $d>4$.

\par
The RG arguements given above imply that
the order parameter obeys the scaling relation
\widetext
\begin{equation}
Q(t,H\Omega, u_1, u_2, . . .) = b^{1-d/2}Q(tb^{1/\nu},H\Omega
b^{1+d/2},u_1b^{4-d}, u_2b^{4-d}, ...)   ,
\label{eq:9}
\end{equation}
\narrowtext
where $\nu = 1/2$, and we have suppressed
all parameters with scale dimensions
smaller than 4-d.  The exponents $\beta$ and z follow from
Eq.\ (\ref{eq:9}) by using standard arguments\cite{MaFisher}.
The crucial point is that $u_1$ and $u_2$ are dangerous irrelevant variables:
Solving the theory explicitly in the saddle point approximation, we see that
$Q(t, \Omega =0, u_1 \rightarrow 0) \sim u_1^{-1/2},$ and $Q(t = 0, \Omega,
u_1
\rightarrow 0, u_2 \rightarrow 0) \sim (u_2/u_1^2)^{1/3}$.
Equation(\ref{eq:9}) then
immediately yields $\beta = 1/2, z = 3$ in agreement with Eq.\ (\ref{eq:1b}).
We emphasize that $K_{s}$ is {\it not} dangerously irrelevant, even though
$Q(t=0,\Omega)$ is proportional to $\sqrt{K_{s}}$. The point is that the
vanishing of the (bare) $K_{s}$ just shifts the transition point to infinity,
making the mean-field transition inaccessible, but does not change the
critical behavior.

\par
We finally derive Eqs.\ (\ref{eq:2}), (\ref{eq:3}).
To this end we notice that all susceptibilities have
been identified in terms of the coupling constants of the NL$\sigma$-
model: $\partial n/\partial {\mu} \sim H+K_s, \chi_s \sim H+K_t,
\gamma \sim H$\cite{Cetal,R}.
$H$, $K_s$, and $K_t$ are all irrelevant, which means that all
three susceptibilities vanish at the critical point.  Furthermore, $K_{s,t}$
scale to
zero faster than H, so the asymptotic scaling behavior of all three
susceptibilities is
the same and given by that of H, or $\gamma$.  The only remaining difficulty
is to correctly incorporate the dangerous irrelevant variables.
Let us consider the singular part $f_{s}$ of the free
energy density $f$, which satisfies the scaling equation\cite{R},
\begin{mathletters}
\label{eqs:10}
\begin{equation}
f_{s}(t,T,u_1,. . .) = b^{-(d+z)}f_{s}(tb^{1/\nu}, Tb^z, u_1b^{4-d}, . . .)  .
\label{eq:10a}
\end{equation}
In the critical regime we have $f_{s} \sim Q^2 \sim 1/u_1$, and hence
\begin{equation}
f_{s}(t,T)=b^{-(4+z)}f_{s}(tb^{1/\nu},Tb^z)  .
\label{eq:10b}
\end{equation}
\end{mathletters}
We see that hyperscaling breaks down in the usual way: $d$ in
Eq.\ (\ref{eq:10a}) gets
replaced by 4.  By differentiating Eq.\ (\ref{eq:10b}) twice with respect to
temperature, and using $z = 3$, we find
\begin{equation}
\gamma (t,T) = b^{-1}\gamma (tb^2,Tb^3)   ,
\label{eq:11}
\end{equation}
and hence Eqs.(\ref{eq:2})\cite{footnote0,footnote1}.
Finally, combination of this result with
Eq.(\ref{eq:Dc}) yields Eq.(\ref{eq:3}). We note that the criticality
of $\partial n/\partial \mu$ is very remarkable, since it does not show
in the $2+\epsilon$ expansion treatment of the same model\cite{F,R},
while it is an important feature of the Mott-Hubbard
transition, which recently has enjoyed a revival of
interest\cite{Hubbardstuff}.

\par
Several questions arise from these results: (1) Can the technical
approach used here be extended to $d<4$? An obvious possibility is to
introduce an (artificial) parameter $N$ multiplying the soft modes, and
perform a $1/N$-expansion in analogy to the $O(N)$-model\cite{ZJ}.
(2) What is the connection between the present results and those obtained
in $d=2+\epsilon$, where $\partial n/\partial \mu$ is found to be uncritical
in perturbation theory\cite{footnote2}?
A possible answer is that $\partial n/\partial \mu$
has an essential singularity near $d=2$ which is invisible in the $2+\epsilon$
expansion. Another possibility is that the Gaussian FP discussed here is not
continuously connected to the one found near $d=2$.
(3) Can the present techniques be extended to the more general
field theory\cite{R} underlying the NL$\sigma$-model? If that is the case,
one would expect the upper critical dimension of the resulting OP description
to be $6$, or, for technical reasons similar to those in
Ref.\onlinecite{HarrisLubensky}, $8$. Which of the two models would capture
more of the physics relevant in $d=3$ would {\it a priori} be unclear.
These problems will be pursued in the future\cite{ustbp}.

\par
In conclusion, we have shown that the generalized NL$\sigma$-model for
interacting disordered electrons, which has been studied extensively in
$d=2+\epsilon$, contains a metal-insulator transition with mean-field
like critical behavior for $d>4$. This transition shows qualitatively
different features from the one obtained in $d=2+\epsilon$: It is order
parameter driven rather than soft-mode driven, and
it shows a critical $\partial n/\partial \mu$.

\par
This work was supported by the NSF under grant numbers DMR-92-17496
and
DMR-92-09879.


\begin{references}
\bibitem{R} For a recent review, see, e.g., D. Belitz and T.~R. Kirkpatrick,
 Rev. Mod. Phys. {\bf 66}, xxx (Apr. 1994).
\bibitem{HarrisLubensky} A.~B. Harris and T.~C. Lubensky, Phys. Rev. B {\bf
23},
 2640 (1981).
\bibitem{WegnerDOS} F. Wegner, Z. Phys. B {\bf 44}, 9 (1981).
\bibitem{FyodorovMirlin} Y.~V. Fyodorov and A.~D. Mirlin, Phys. Rev. Lett.
 {\bf 67}, 2049 (1991).
\bibitem{F} A.~M. Finkel'stein, Zh. Eksp. Teor. Fiz. {\bf 84}, 168 (1983)
 [Sov. Phys. JETP {\bf 57}, 97 (1983)];
 this is a generalization of Wegner's model for
 noninteracting electrons, F. Wegner, Z. Phys. B {\bf 35}, 207 (1979).
\bibitem{ZJ} See, e.g., J. Zinn-Justin, {\it Quantum Field Theory and
 Critical Phenomena} (Clarendon, Oxford 1989).
\bibitem{ustbp} D. Belitz and T.~R. Kirkpatrick, unpublished.
\bibitem{AA} B.~L. Altshuler and A.~G. Aronov, Solid State Commun. {\bf 30},
 115 (1979).
\bibitem{MaFisher} S.-K. Ma, {\it Modern Theory of Critical Phenomena}
 (Benjamin, Reading, MA 1976); M.~E. Fisher, in {\it Advanced Course on
 Critical Phenomena}, edited by F.~W. Hahne (Springer, Berlin 1983), p.1.
\bibitem{Cetal} C. Castellani, C. DiCastro, P.~A. Lee, M. Ma, S. Sorella,
 and E. Tabet, Phys. Rev. B {\bf 33}, 6169 (1986).
\bibitem{Hubbardstuff} For a recent review, see, D. Vollhardt in
 {\it Correlated Electron Systems}, edited by V. Emery (World Scientific,
 Singapore, 1993).
\bibitem{footnote0} We note that Eqs.(\ref{eqs:10}) by themselves give
 information only about the {\it singular} part of $\gamma$. However,
 taken together with the knowledge that $H$ scales to zero they yield
 Eqs.(\ref{eq:2}).
\bibitem{footnote1} The Eqs.(\ref{eq:2}) also follow from the
 following argument. Except for logarithmic corrections, the hyperscaling
 relation, $\sigma \sim \xi^{-(d-2)}$, is expected to be valid up to the
 upper critical dimension, i.e., $d=4$. Using $D_{c} \sim \xi^{-1}$ and
 the Einstein relation, $\sigma = D_{c} \partial n/\partial \mu$, gives
 $\partial n/\partial \mu \sim \chi \sim \xi^{-1}$, i.e. Eqs.(\ref{eq:2}).
 From general arguments concerning the breakdown of hyperscaling one
 expects this result to hold for $d>4$ as well.
\bibitem{footnote2} In $d=2$, one does not expect logarithmic corrections
 to $\partial n/\partial \mu$ to all orders in perturbation theory. The
 reason is that $\partial n/\partial \mu$ is given as a frequency integral
 over a quantity which itself is only logarithmically singular.
\end{references}
\end{document}